\documentclass[10pt,conference]{IEEEtran}
\IEEEoverridecommandlockouts

\usepackage{amsmath,amssymb,amsfonts}
\usepackage{graphicx}
\usepackage{textcomp}
\usepackage{lipsum}
\usepackage{tcolorbox}
\usepackage{cite}
\usepackage{amssymb}
\usepackage{soul}
\usepackage{multirow}
\usepackage{colortbl}
\usepackage{graphicx}
\usepackage{textcomp}
\usepackage{ifpdf}
\usepackage{enumitem}
\usepackage{varwidth}
\usepackage{authblk}
\usepackage{amsmath}
\usepackage[framemethod=tikz]{mdframed} 
\usepackage{microtype}
\usepackage{color}
\usepackage{tabularray}
\definecolor{Alto}{rgb}{0.839,0.839,0.839}
\definecolor{Mercury}{rgb}{0.901,0.898,0.898}
\usepackage{url}

\newcommand{\TM }{Task-Focused Mentoring Strategies} 
\newcommand{\SM }{Mentor Attributes and Ideal Outcomes}

\def\BibTeX{{\rm B\kern-.05em{\sc i\kern-.025em b}\kern-.08em
    T\kern-.1667em\lower.7ex\hbox{E}\kern-.125emX}}

\usepackage{multibib}
\newcites{lit}{References B: Conceptual Framework of \SM{}}

\newtcolorbox{bubble}[1][]{colback=blue!5!white, colframe=blue!75!black,
  fonttitle=\bfseries, title=#1,  width=\linewidth, 
  boxsep=1pt, top=1pt, bottom=1pt}

\let\svthefootnote\thefootnote
\newcommand\blankfootnote[1]{%
  \let\thefootnote\relax\footnotetext{#1}%
  \let\thefootnote\svthefootnote%
}

\title{The Multifaceted Nature of Mentoring in OSS: Strategies, Qualities, and Ideal Outcomes }

\author{\small%
    Zixuan Feng\textsuperscript{1}, 
    Igor Steinmacher\textsuperscript{2}, 
    Marco Gerosa\textsuperscript{2}, 
    Tyler Menezes\textsuperscript{3}, 
    Alexander Serebrenik\textsuperscript{4}, 
    Reed Milewicz\textsuperscript{5}, 
    Anita Sarma\textsuperscript{1}\\
    \textsuperscript{1}Oregon State University, USA, \{fengzi, anita\}@oregonstate.edu;\\
    \textsuperscript{2}Northern Arizona University, USA, \{marco.gerosa, igor.steinmacher\}@nau.edu;\\
    \textsuperscript{3}CodeDay, USA, tylermenezes@codeday.org;\\
    \textsuperscript{4}Eindhoven University of Technology, Netherlands, a.serebrenik@tue.nl;\\
    \textsuperscript{5}Sandia National Laboratories, USA, rmilewi@sandia.gov
}

\begin{document}

\maketitle

\begin{abstract}
Mentorship in open source software (OSS) is a vital, multifaceted process that includes onboarding newcomers, fostering skill development, and enhancing community building.
This study examines task-focused mentoring strategies that help mentees complete their tasks and the ideal personal qualities and outcomes of good mentorship in OSS communities. 
We conducted two surveys to gather contributor perceptions: the first survey, with 70 mentors, mapped 17 mentoring challenges to 21 strategies that help support mentees. The second survey, with 85 contributors, assessed the importance of personal qualities and ideal mentorship outcomes. 
Our findings not only provide actionable strategies to help mentees overcome challenges and become successful contributors but also guide current and future mentors and OSS communities in understanding the personal qualities that are the cornerstone of good mentorship and the outcomes that mentor-mentee pairs should aspire to achieve.
\end{abstract}


\section{Introduction}

It is well-recognized that mentorship in software engineering, as in the broader workforce, plays a critical role in shaping both the profession and those within it. From an individual perspective, Ragins and Kram, 
authors of the \textit{Handbook of Mentoring at Work} observe, ``When asked to contemplate relationships that make a difference in our lives---relationships that have given us the courage we think we cannot do, relationships that have guided our professional development or even changed the course of our lives---many of us think of mentoring relationships''\cite{ragins2007handbook}. Moreover, mentorship helps develop and sustain professional communities; taken together, mentor-mentee relationships can create resilient networks of mutual support and can enable the growth and maintenance of communities of practice\cite{tarr2010tangled, rinne2023give, feng2022case}.

These dual aspects of mentorship are especially important in the context of OSS projects and their sustainability, where mentorship not only trains new and current contributors in improving technical skills but also helps them to navigate and become part of the community's culture and social dynamics \cite{steinmacher2016overcoming, feng2024guiding}. These, among other reasons, are why many organizations formally support mentoring programs, such as the Linux Foundation mentoring programs \cite{linux_mentorship}, Google Summer of Code \cite{gsoc2024}, and CodeDay \cite{menezes2022open}, where newcomers are paired with mentors who provide technical guidance and community support. Given the potential of mentorship to make OSS projects more enduring and productive, from a research perspective, we are interested in what makes good mentor-mentee relationships and what kinds of strategies could make them more successful.

In practice, the primary focus of OSS mentorship programs is on securing newcomers and matching them with mentors. Less attention is paid to the performance of mentorship by mentors, leaving individual mentors to determine best practices for mentorship on their own. Mentors are expected to take on this role in addition to their regular responsibilities and day jobs \cite{balali2018newcomers}. In addition to a lack of time and resources, mentors often face other mentoring challenges as documented by numerous studies \cite{balali2020recommending,steinmacher2016overcoming, steinmacher2019overcoming ,Jacobs2024}; we direct readers to a recent systematic literature review by Feng et al. for a comprehensive overview of this topic\cite{feng2024guiding}. As an example, mentors commonly face challenges in understanding the diverse backgrounds of mentees, which often leads to difficulties in identifying tasks that not only support mentees' skill development but also align with mentees' interests and expectations. Gaining that depth of understanding typically requires a level of trust that can be more difficult to establish in virtual settings\cite{feng2024guiding}. This, in turn, raises questions about how mentors might engender that trust and what kinds of practical strategies they should employ to help the mentee achieve their desired outcomes. That is, we see a throughline between mentor behaviors and characteristics (who mentors should be to their mentees), strategies for supporting mentees (what mentors should do), and outcomes of mentorship (why mentors should do those things). 

In the literature on OSS mentorship, there has been relatively little attention paid to connecting these ideas together in a coherent way. To the best of our knowledge, Balali et al.\cite{balali2020recommending}, is the only work that, through a survey of 30 contributors, mapped strategies that mentors follow to overcome challenges in finding the right task to recommend to newcomers. The goal of our work is to build a more comprehensive mapping (beyond just task recommendation) that threads the needle on effective strategies for OSS mentorship. To that end, we developed three research questions to guide our study:

\begin{itemize}
\setlength{\itemindent}{1em}
\item[\textbf{RQ1}] What are the strategies mentors perceive as effective to address challenges mentees face in contributing to OSS?
\item[\textbf{RQ2}] What are the qualities of an ideal mentor?
\item[\textbf{RQ3}] What are ideal outcomes of good mentorship?
\end{itemize}

To answer \textbf{RQ1}, we mapped challenges that mentors face and potential strategies as collected from Feng et al.'s literature review \cite{feng2024guiding}, which we then combined with feedback from OSS mentors from the Outreachy leadership team \cite{outreachy} and the head of Linux Foundation mentoring program \cite{linux_mentorship}. We then surveyed 70 mentors who mapped the 21 challenges to the 17 strategies we identified (For the purposes of this paper, we refer to these as \TM{}).

But, mentoring is a multifaceted process that goes beyond technical skill building and the ability to complete project tasks. It also encompasses a broader and deeper range of activities to nurture the long-term development of an individual \cite{feng2022case, gallacher1997supervision}. 
Understanding the ideal personal characteristics of a mentor and the ideal outcomes of mentoring is important to be able to build the right mentor-mentee relationship.

We answer \textbf{RQ2} and \textbf{RQ3} by first identifying the personal qualities of mentors and ideal mentorship outcomes through a review of 57 mentoring papers. We then conducted a survey of 85 contributors to assess the importance that contributors perceive about these attributes of an ideal mentor and outcomes of good mentorship; we refer to these as \SM{} in the paper.

The overarching goal of our work is to lay out what matters in OSS mentorship. We hope that the answers to these research questions will not only provide mentors with actionable strategies to guide their mentees to become successful OSS contributors, but also shed light on what makes for good OSS mentors and the diverse range of mentorship outcomes.

\label{sec:introduction}
\vspace{-2mm}

\section{Related Work}

Mentoring in OSS is a multifaceted process that includes onboarding of newcomers, as well as continued growth and development that is often achieved through peer mentoring and community engagement \cite{feng2022case, feng2024guiding, steinmacher2019overcoming, balali2018newcomers}. This section focuses on the ``traditional" mentoring aspects of onboarding.

\textbf{\TM{}}: Onboarding newcomers to an OSS project by helping them make their first contribution is essential, as newcomers need guidance on the technical aspects and the project’s processes and culture. Much of the existing research on OSS mentoring has focused on understanding the challenges mentors face and proposing task-focused strategies to address them \cite{feng2024guiding, balali2018newcomers, balali2020recommending, steinmacher2019overcoming}. For example, strategies such as suggesting small, interest-aligned tasks, tagging task complexity, and maintaining documentation have been suggested to mitigate challenges like skill gaps \cite{steinmacher2019overcoming}. However, it is not known which strategies are effective in overcoming which challenge. We are aware of only one study that investigated the effectiveness of strategies that mitigate challenges in recommending tasks to newcomers in OSS \cite{balali2020recommending}. 
However, many more mentoring challenges exist; Feng et al. \cite{feng2024guiding} conducted a systematic literature review that developed a taxonomy of challenges and potential strategies in OSS mentoring. This gap in our knowledge of which strategies work for which challenges could potentially lead to a disconnect between strategies and contributors' actual needs or preferences, thereby reducing the effectiveness of mentorship initiatives.

\textbf{\SM{}} Additionally, interpersonal characteristics also play a key role in mentoring. Psychological and emotional support are positively associated with a sense of belonging among contributors \cite{trinkenreich2023belong}. Psychological support includes reassurance, encouragement, constructive feedback, and guidance in navigating expectations, while emotional support involves empathy and acknowledgment of individuals' efforts \cite{ragins2007handbook, verenikina2008scaffolding}. Both psychological and emotional support are essential in OSS, where they contribute to a supportive environment and community engagement. Trinkenreich et al. \cite{trinkenreich2023belong} found that intrinsic motivations, such as social or hedonic motives, are positively associated with a sense of virtual community, which can be fostered through effective psychological and emotional support. However, few studies in OSS mentoring explore how these interpersonal characteristics are perceived as a requirement for good mentorship.

\textbf{Our contribution} Our study fills two gaps: (1) the limited understanding of actionable strategies that mentors perceive as effective in supporting mentees become successful contributors, and (2) the lack of understanding of what is perceived as good qualities of a mentor and the ideal outcome of mentoring.

\label{sec_related}
\vspace{-2mm}

\section{Survey Design}

To answer our research questions, we conducted two surveys: 1) \TM{}, which focuses on gaining insights into effective strategies for mentors to support mentees in successfully completing their tasks. (2) \SM{}, which focuses on perceived interpersonal characteristics, including the personal attributes of an ideal mentor and the ideal outcomes of good mentorship. We chose to conduct two surveys rather than a single large one to avoid participant fatigue due to survey length (which would have exceeded 40 minutes) and to increase the likelihood of receiving responses \cite{sahlqvist2011effect}. The following sections outline the study design of each survey.  The survey questions used for the study are available in the companion website \cite{anonymous}.

\subsection{\TM{}}

\textbf{Survey Overview}: The \TM{} survey comprised four sections, the first section is a consent form and a screening question asking whether the participant had served as a mentor in any capacity within OSS. Since the strategies are intended for mentors, participants who answered ``no" to the screening question automatically exited the survey. The second section focused on mentor demographics, including their OSS seniority, mentorship experience, and identity.

In Section 3 of the survey, we asked participants to map strategies to challenges by selecting one or more strategies that effectively address each challenge. The strategies and challenges listed in the survey were based on the framework developed by Feng et al. \cite{feng2024guiding}. A matrix was provided for this set of questions. We divided the large mapping task into three sub-questions based on the categories of challenges identified by Feng et al.\cite{feng2024guiding} (social-related challenges, process-related challenges, and technical-related challenges)

The survey concluded with an open-ended question asking participants if there were anything else they would like to add about the survey or mentoring. We obtained approval from the university's Institutional Review Board (IRB), and participants were informed that all questions were optional and that they could contact us with any concerns. To encourage participation, we held \$50 Amazon gift card raffles for every 100 participants.

\textbf{Pilot Survey}: Before distributing the survey, we conducted three rounds of pilot. In the first round, we tested the survey with five graduate students with software-related research experience. Based on their feedback, we split the three matrix mapping questions into five subsections for easier comprehension. The participants noted that longer questions made it difficult to map the strategies to the challenges.

In the second round, we piloted the survey with the leadership team of Outreachy \cite{outreachy}, who also have mentoring experience. In this pilot survey, the mapping presented 13 challenges and 17 strategies as synthesized in \cite{feng2024guiding}. Participants suggested adding eight more challenges by refining some of these challenges into more specific ones. For example, \textit{unsupportive environment} in \cite{feng2024guiding} was defined as ``\textit{Mentees face difficulties with a hostile project environment, slow response rates, and securing assistance},''. This was divided into two challenges: \textit{delayed assistance} and \textit{hostile environment}. Similarly, the challenge \textit{low self-efficacy} was split into \textit{fear of feedback} and \textit{resistance to guidance} to improve clarity and targeted responses. In total, there were 17 strategies to map to 21 challenges (See supplementary \cite{anonymous}). Lastly, we conducted a third pilot round with the leaders of the Linux mentoring program; they have extensive experience as mentors. In this round, only minor wording suggestions were made.

\textbf{Data Collection}: We advertised the survey through three approaches and received 115  responses. 1) We advertised the survey at two OSS conferences, which garnered 23 responses. 2) We then promoted the survey on social media, including Twitter and LinkedIn (n=17). 3) We then asked the president of the CodeDay \cite{menezes2022open}, to advertise the survey, resulting in 72 responses. We used Qualtrics \cite{qualtrics} to deploy our survey. 17 participants didn't pass the screening question (not mentor).

After removing 28 incomplete answers, 70 responses remained. Of these, 42 participants (60.0\%) had been mentors for less than 5 years, while 28 (40.0\%) had been mentors for 5 years or more. 15 participants (21.4\%) identified as women contributors, and 5 identified as having disabilities. 56 (80.0\%) were from North America, 5 (7.1\%) from Europe, 5 (7.1\%) from Asia, 3 (4.3\%) from South America, and 1 participant preferred not to disclose their region.

\subsection{\SM{}}

\textbf{Conceptual Framework for \SM{}} Effective mentorship go beyond task-oriented support~\cite{kram1985mentoring}, the focus of \SM{} is to explore mentorship attributes through a broader lens. We grounded our conceptual framework of mentorship attributes using Kram's mentor role theory~\cite{kram1985mentoring}. This framework defines mentorship as serving two main functions: career development and psychosocial support. Career mentoring helps mentees develop the competencies they need to thrive in the workplace and mature as professionals. Psychosocial mentoring focuses on supporting the mentee's personal development and sense of belonging in their career and organization. However, mentoring in OSS differs from conventional organizational settings. OSS is volunteer-led, with contributors collaborating globally in public spaces and often relying on asynchronous communication channels \cite{feng2022case,feng2024guiding}. These dynamics may change how mentors provide career and psychosocial support. Using Kram's model as a starting point, we identified two conceptual categories for our framework: the ideal qualities of mentors (\textit{e.g.,} being trustworthy) and the ideal outcomes of mentorship.

To populate our conceptual framework, we reviewed existing mentoring literature, including sources beyond OSS and SE. This approach allowed us to build a framework that applies to OSS but is also broad enough to allow meaningful comparisons with mentorship practices in other fields. Two researchers reviewed related literature in Google Scholar. Our final keywords were inspired by Kram's mentor role theory~\cite{kram1985mentoring}, such as ``mentoring," ``mentor," ``mentorship," ``characteristics," ``qualities," and ``outcomes." The initial search yielded 995 studies. Then, two authors read the titles and abstracts, and only selected papers discussed mentoring in the abstract (N=479).

Two researchers independently assessed accessibility, using institutional resources to download papers, resulting in 436 accessible studies. To focus on the robustness and credibility of the papers, we removed those without citations, which resulted in 354 papers. Next, two researchers independently evaluated each study for relevance to mentoring characteristics, qualities, and outcomes, undergoing three rounds of inter-rater reliability assessments to resolve discrepancies. This process culminated in a final set of 57 papers that discussed attributes of good mentorship across different disciplines, listed in Reference B.

We then read these papers and performed thematic analysis \cite{kaur2024challenges, braun2012thematic} and open coding \cite{howard2016glaser} to identify different qualities of mentors and outcomes of mentorship that are considered ideal. Two researchers met weekly to discuss the emerging code and develop the preliminary themes, continuously review and update the codes, and resolve disagreements through negotiated agreement \cite{garrison2006revisiting}. We continued coding until no new themes emerged. In total, we identified a set of 9 qualities of good mentors and 11 mentorship outcomes (Table \ref{tab:framework}).

\begin{table*}
\centering
\caption{A list of attributes used in the conceptual framework of software engineering mentorship. definitions are provided for frequently mentioned and highly ranked attributes.}
\label{tab:framework}
\resizebox{\textwidth}{!}{
\begin{tblr}{
  row{3} = {Mercury},
  row{5} = {Mercury},
  row{7} = {Mercury},
  row{9} = {Mercury},
  row{11} = {Mercury},
  cell{1}{1} = {c=3}{Mercury,c},
  cell{1}{4} = {r=12}{},
  cell{1}{5} = {c=3}{Mercury,c},
}
\textbf{Qualities of an Ideal Mentor} &                                                                                           &                                                                                                                                                  &  & \textbf{Ideal Outcomes of Mentorship}                            &                                                                                                                                                                &                                                                                                                                                                                                                                                                                                                                                       \\
\textbf{Nonjudgmental}                & {Patient, accepting, open to \\mentee’s ideas, and does \\not rush to judge the mentee.}  & {\citelit{paper54,alex182,paper480, paper650,alex148}\\\citelit{paper408,paper532}}                                                              &  & {\textbf{Coaching and }\\\textbf{Vision-Building}}               & {Mentor advises on career \\progress, including \\achievement of appropriate \\career milestones.}                                                             & {\citelit{paper47,paper152,paper154,paper480,paper513}\\\citelit{paper571,paper650,paper781,paper21,paper26}\\\citelit{alex124,alex240,paper301,alex553,paper614}\\\citelit{paper625,alex726,alex765,alex849,paper914}}                                                                                                                               \\
\textbf{Active Listener}              & {Has demonstrated \\good listening skills \\in conversations.}                            & {\citelit{paper54, paper154, paper182, paper513}\\\citelit{paper236}}                                                                            &  & {\textbf{Encouraging }\\\textbf{Skill }\\\textbf{Development}}   & {Mentor helps mentee develop \\skills needed in their career~\\with the end the goal of making \\the mentee more self-sufficient, \\independent, and capable.} & {\citelit{paper28,paper47,paper54,paper55,paper152}\\\citelit{paper153,paper182,alex182,alex304,paper480}\\\citelit{paper513,paper570,paper571,paper572,paper650}\\\citelit{alex717,paper787,paper21,paper26,paper59}\\\citelit{alex64,paper197,alex553,paper614,paper625}\\\citelit{paper642,paper668,alex668,alex726,paper858}\\\citelit{paper859}} \\
\textbf{Empathetic}                   & {Care and conveys \\empathy for the \\concerns and feelings \\discussed by the \\mentee.} & {\citelit{paper54,paper154,alex182,paper943,paper21}\\\citelit{paper112,paper408,paper989}}                                                      &  & {\textbf{Promoting Self-}\\\textbf{Awareness}}                   & {Mentor uncovers mentee's \\underlying assumptions \\through careful probing \\and scaffolds them into \\deeper levels of thinking.}                           & {\citelit{paper47,paper54,paper55,paper152,paper182}\\\citelit{alex182,alex765,paper570,paper650,alex717}\\\citelit{paper26,alex64,paper112,paper197,paper236}\\\citelit{alex240,paper408,paper532,paper914,paper614}\\\citelit{alex668,alex726}}                                                                                     \\
\textbf{Trustworthy}                  & {Inspires and earns \\the trust of the \\mentee.}                                         & {\citelit{paper54,paper154,paper182,alex182,paper858}\\\citelit{paper112,alex309,alex367,paper408,paper532}\\\citelit{alex555,paper668,alex849}} &  & {\textbf{Providing Emotional }\\\textbf{and Moral Support}}      & {Mentor helps the mentee to \\clarify feelings, permits \\vulnerability encourages \\discussion of the personal \\meaning of experiences.}                     & {\citelit{paper54,paper55,paper152,paper154,alex304}\\\citelit{paper480,paper570,paper571,paper781,paper26}\\\citelit{alex124,paper301,alex553,paper614,paper625}\\\citelit{paper642,alex668,alex726,alex765,paper888}\\\citelit{paper914,paper989}}                                                                                                  \\
\textbf{Honest}                       & {Genuine, open, willing \\to explain themselves, \\and holds themselves \\accountable.}   & {\citelit{paper54,paper154,paper182,paper943,paper21}\\\citelit{paper236,paper367}}                                                              &  & {\textbf{Navigating the }\\\textbf{Institution}}                 & {Mentor helps them to adapt \\to the norms, standards, and \\expectations associated with\\their profession.}                                                  & {\citelit{paper54,paper55,paper152,paper153,paper154}\\\citelit{alex182,paper480,paper513,paper571,paper572}\\\citelit{paper943,alex64,paper197,alex726,paper642}}                                                                                                                                                                 \\
\textbf{Encouraging}                  & {Supportive and motivates \\the mentee to do their \\best work.}                          & {\citelit{paper54,paper55,paper182,alex182,alex502}\\\citelit{paper21,paper236,paper532,paper614,alex765}\\\citelit{paper989}}                   &  & \textbf{Cultivating Networks}                                    & {Mentor helps mentee gain \\access to otherwise closed \\circles, increase their contact \\with people~for future.}                                            & {\citelit{paper54,paper152,paper154,alex182,paper513}\\\citelit{paper570,paper571,paper572,paper21,paper26}\\\citelit{alex64,alex124,paper197,paper532,paper625}\\\citelit{paper642,alex726,alex849}}                                                                                                                                                 \\
\textbf{Respectful}                   & {Conveys feelings of \\respect for the mentee \\as an individual.}                        & {\citelit{paper54,paper154,paper182,alex182,paper650}\\\citelit{paper943,paper21,paper112,paper408,alex849}}                                     &  & \textbf{Acting as Role Model}                                    & {Mentor provides the mentee \\with a model for what their \\future self will look like, \\mentor when they reach a \\similar position in their career.~}       & {\citelit{paper54,paper152,paper182,paper480,paper570}\\\citelit{paper571,alex717,paper781,paper915,paper943}\\\citelit{paper21,alex64,paper112,paper236,alex148}\\\citelit{alex240,paper301,paper408,alex553,alex726}\\\citelit{alex765,alex849,paper989}}                                                                                           \\
\textbf{Approachable}                 & {Enthusiastic, friendly, \\and a good sense \\of humor}                                   & {\citelit{paper54,paper55,paper154,paper781,paper943}\\\citelit{paper21,paper112,alex148,paper408,alex555}\\\citelit{paper614,alex726,paper989}} &  & {\textbf{Protecting and }\\\textbf{Advocating}}                  & {Mentor advocates for mentee \\in the team while protecting\\them from harsh interactions.}                                                                    & {\citelit{paper54,paper154,paper182,alex182,paper480}\\\citelit{paper571,alex553,paper625,paper859,paper914}\\\citelit{paper989}}                                                                                                                                                                                                                     \\
\textbf{Reliable}                     & {Responsive and does \\what they say they'll do.}                                         & \citelit{paper54, paper642,paper858}                                                                                                           &  & \textbf{Offering Friendship}                                     & {Mentor interacts with mentee \\socially outside of work.}                                                                                                     & {\citelit{paper54,paper55,paper513,paper570,paper408}\\\citelit{alex553,paper914}}                                                                                                                                                                                                                                                                    \\
                                      &                                                                                           &                                                                                                                                                  &  & {\textbf{Instilling a Sense of }\\\textbf{Satisfication}}       & {Mentor feels an intrinsic, \\personal satisfaction.}                                                                                                          & {\citelit{paper47,paper54,paper5,paper154,paper513}\\\citelit{paper570,paper571,paper787,paper532,paper642}\\\citelit{alex765}}                                                                                                                                                                                                                       \\
                                      &                                                                                           &                                                                                                                                                  &  & {\textbf{Reinforcing Mentors' }\\\textbf{Professional Identity}} & {Mentoring reinforces mentors' \\professional identity, status, \\and self-worth}                                                                              & {\citelit{paper47,paper54,paper55,paper570,paper532}\\\citelit{alex765}}                                                                                                                                                                                                                                                                              
\end{tblr}}
\vspace{-5mm}
\end{table*}

\textbf{Survey Overview} The \SM{} focused on philosophies of effective mentoring, which is often universal and transferable between SE and OSS~\cite{kram1985mentoring}. Therefore, unlike the \TM{}, which focused on contribution challenges in OSS and distributed to OSS mentors, the \SM{} survey included participants from the wider SE field, including those from OSS. 
Additionally, as we were investigating the reciprocal relationship in mentorship, we collected the perceptions of both mentors and mentees about what they thought were the ideal mentor quality and ideal outcome of mentorship.

The \SM{} survey consisted of four sections. The first section included a consent form and a screening question to determine if participants had ever been mentored or mentored by someone else.  In the second section, for the qualities of ideal mentors and the expected outcomes of mentorship, participants were asked to rank each characteristic on a Kano scale \cite{witell2013theory}: which helps distinguish between (1) essential (Must-have), (2) Good to Have (attractive), (3) Unimportant, or (4) Harmful (undesirable) characteristics, and (5) I Don’t Know (uncertainty). This scale has been successfully used in software engineering research \cite{begel2014analyze}. These two sets of questions helped us to identify which aspects can be strategic in fostering effective mentor-mentee relationships.  The survey questions were approved by IRB. Participants were informed that all questions were optional, that they were not required to answer any question that made them uncomfortable, and that they could contact us with any questions or comments.

\textbf{Data Collection}: Before we began the survey, we piloted the questions, which helped us to clarify the instructions and questionnaire language. We advertised the survey through multiple channels to recruit participants. We began by promoting the survey on social media, posting on Facebook, X/Twitter, LinkedIn, Mastodon, and Reddit, and we contacted developers within the US-RSE (US Research Software Engineer Association) via their Slack channel. Meanwhile, following the principle of social benefit, we pledged to donate \$1 USD to an open-source foundation or charity for every completed survey, allowing participants to choose which foundation their dollar would go to. Additionally, to encourage participation, as part of the US-RSE slack and social media recruitment, we held two \$100 Amazon gift card raffles. After removing unfinished responses, we received a total of 85 valid responses.

Among the participants, 58 had experience as both a mentor and mentee, 11 had only been mentors, and 11 had only been mentees. Regarding gender, 18 participants (21.2\%) identified as women, 59 (69.4\%) identified as men, 3 (3.5\%) identified as non-binary, and 5 preferred not to disclose. Of the 85 participants, 56 (65.9\%) are from North America, 24 (28.2\%) from Europe, 2 (2.4\%) from Asia, 2 (2.4\%) from South America, and 1 (1.2\%) from Africa.

\subsection{Data analysis:} We quantitatively analyzed both survey response datasets. To categorize the challenges in the \TM{} survey, we restructured 21 identified challenges into six categories by combining insights from an existing study~\cite{feng2024guiding} and feedback from pilot study participants. This reorganization was done to clarify the broader themes and a cleaner presentation of the data analysis. The six categories are (1) lack of skills, (2) communication barriers, (3) unclear mentees' expectations and background, (4) project climate, (5) lack of resources, and (6) project organization (See supplementary \cite{anonymous} for challenge and strategy categories).

\label{sec_method}
\vspace{-2mm}

\section{Results}

Section \ref{sec:result1} describes the \TM{} frequently used to manage specific mentoring challenges in helping mentees contribute to OSS. In Section \ref{sec:result2}, we detail the mentor attributes and mentorship outcomes that are considered ideal. Together, our findings offer mentors a systematic guide they can use to help mentees navigate the social and technical demands of OSS projects.

\subsection{\TM}
\label{sec:result1}

\begin{figure*}
\centering
\includegraphics[width=5.5in]{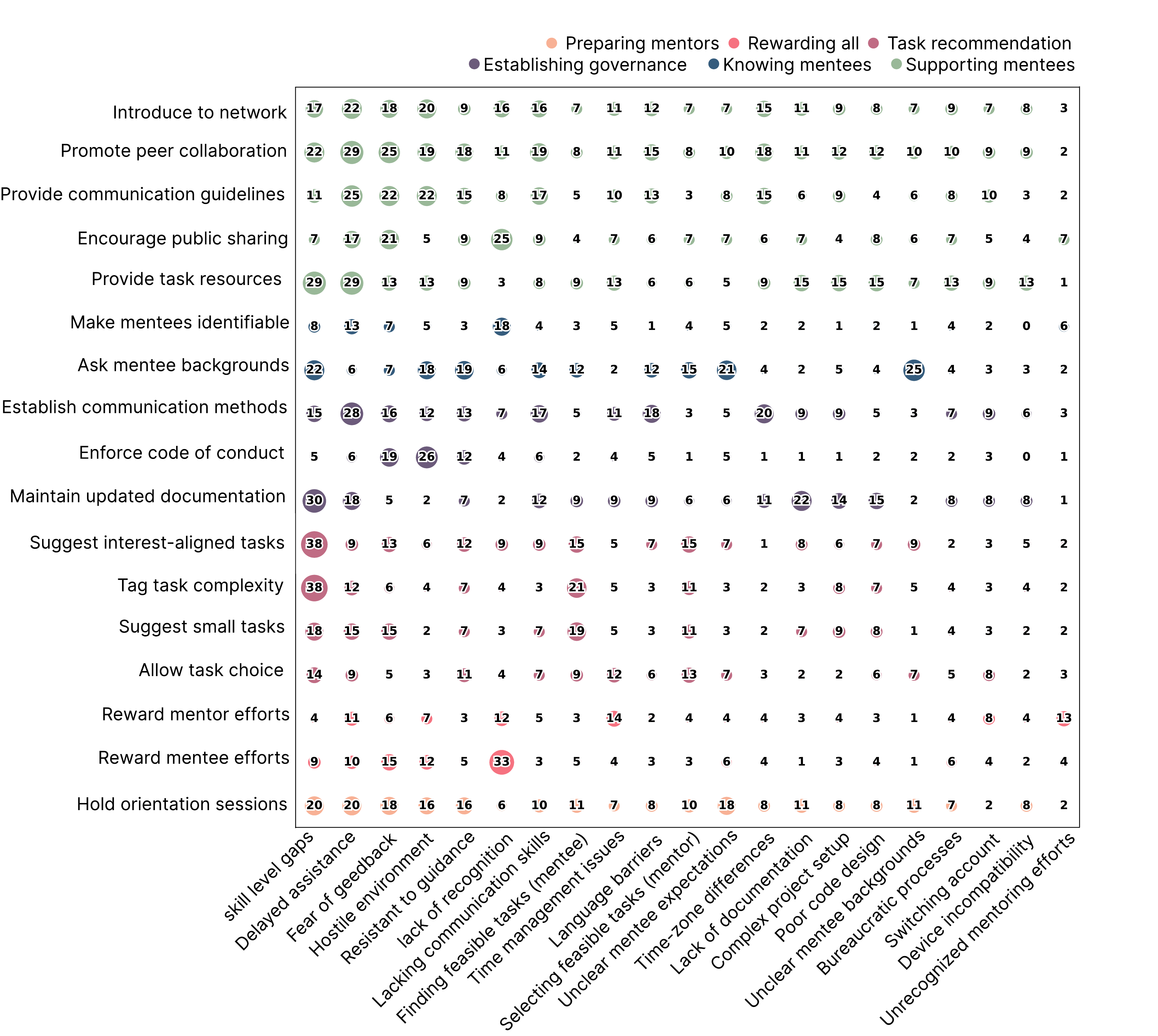}
\caption{The vertical axis represents strategies, while the horizontal axis represents challenges. Each challenge could be linked to multiple strategies for mitigation. The size of each bubble reflects the number of responses mentors provided for a particular strategy mitigating a specific challenge, and the color coding indicates different categories of strategies as explained in Section \ref{sec_method}. Figures \ref{fig:preparing} to \ref{fig:supporting_mentees} provide a zoomed-in analysis of each strategy category.} 
\label{fig:overview}
\vspace{-5mm}
\end{figure*}

Mentoring is not easy, and several studies have reported the plethora of challenges that mentors face \cite{feng2024guiding}, such as unclear mentee expectations, skill gaps, balancing time and responsibilities, and more. Here, we map the strategies our survey participants (mentors) perceived as effective in addressing their challenges.

Figure~\ref{fig:overview} shows an overview of the \TM{} mapped to the challenges. The left-most column shows the challenge, e.g.,  \ul{skill level gaps}, which can be addressed by several strategies (rows in Figure \ref{fig:overview}); the top three strategies to help with this challenge are: \ul{suggest interest-aligned tasks} (38 responses), \ul{tag task complexity} (38 responses), and \ul{maintain updated documentation} (30 responses). The first of these three strategies tries to minimize the gap by finding tasks suited to the mentees' interests, while the latter two make it easier to understand the task and the project. The visualization provides a bird’s-eye view of ``OSS-mentor-vetted'' strategies mapped to challenges, which mentors can adopt as suited to their project organization and available resources.

\begin{figure}[!tbp]
\centering
\includegraphics[width=3.4 in]{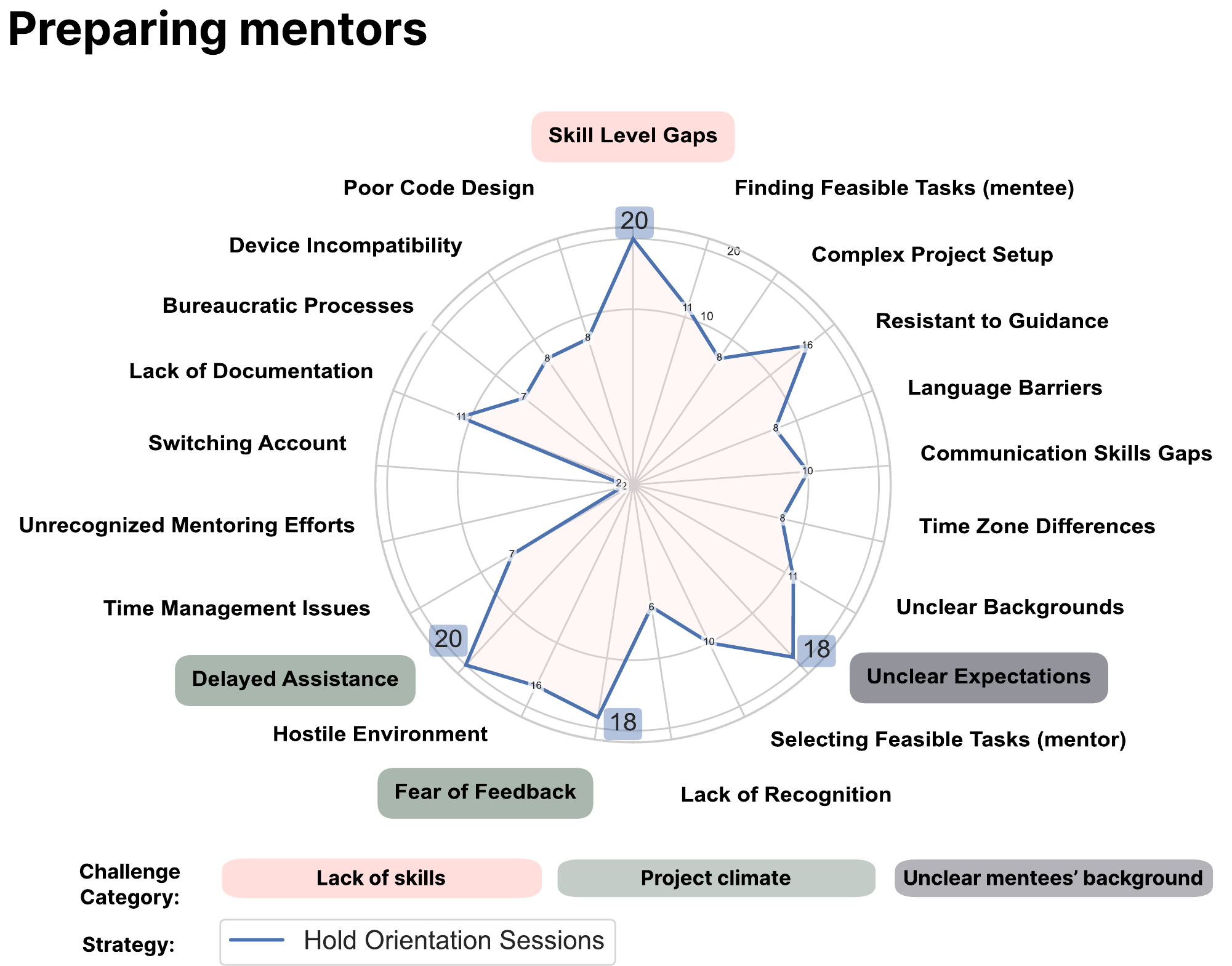}
\caption{Mapping of challenges to the strategy category \textbf{preparing mentors}. The polygon axis points depict the survey participant responses of challenges that are mitigated by the strategy  \ul{hold orientation sessions}. Only challenges in the top 90th percentile of responses are shown, color-coded by their challenge category: lack of skills (pink), project climate (dark green), and lack of resources (gray). Subsequent figures follow the same layout.} 
\label{fig:preparing}
\vspace{-4mm}
\end{figure}

We here present the survey results disaggregated for each of the six strategy categories \cite{feng2024guiding} (Figures \ref{fig:preparing} to \ref{fig:supporting_mentees}). Each of the strategies in a category is presented through a spider plot, which we chose to be able to visually compare multiple variables (challenges) simultaneously. In the plot, each challenge is represented as a spoke (or axis) radiating from a central point, and the number of respondents who selected a particular strategy for that challenge is plotted along these axes. These values are then connected to form a polygonal shape, making it easy to identify patterns and determine which strategy is most effective for addressing specific challenges.

For instance, as shown in Figure \ref{fig:preparing}, holding orientation sessions (represented by the blue polygon) effectively addresses challenges such as skill level gaps and delayed assistance, each receiving 20 responses. Similarly, it is also helpful for fear of feedback and unclear expectations, both with 18 responses. The responses are marked on the spider plot using the same color as the strategy line (here, blue). To reduce visual clutter, we only display responses that fall within the top 90th percentile.

Sometimes, a strategy category (e.g., Figure \ref{fig:task}) can have multiple strategies. In such cases, each strategy is depicted using different colored lines.  
Additionally, each challenge is categorized and color-coded (as shown at the bottom of each figure), allowing for easy mapping of highest-responsed challenges in their respective categories that are mitigated by that strategy. In the rest of the section, we only discuss these top-responsed challenges when discussing the strategies.

\subsubsection{Preparing Mentors} includes only one strategy \ul{holding orientation sessions}:

\textbf{Orientation for Lowering Barriers.} Orientation sessions help set clear expectations and introduce mentees to project norms, coding standards, and available resources \cite{balali2018newcomers}. Such sessions can also set guidelines and articulate mentor responsibilities \cite{bland2009faculty}.  Figure \ref{fig:preparing} shows that a majority of mentors found orientation sessions (blue highlighted numbers indicating \#responses) to address the challenge category regarding lack of skills, unclear mentees' background, project climate.

\begin{bubble}[Orientation sessions...]
support mentees in skill-building, task comprehension, understanding expectations, and fostering confidence in the mentoring process.
\end{bubble}

\subsubsection{Rewarding Mentees and Mentors} 
\begin{figure}[!tbp]
\centering
\includegraphics[width=3.4 in]{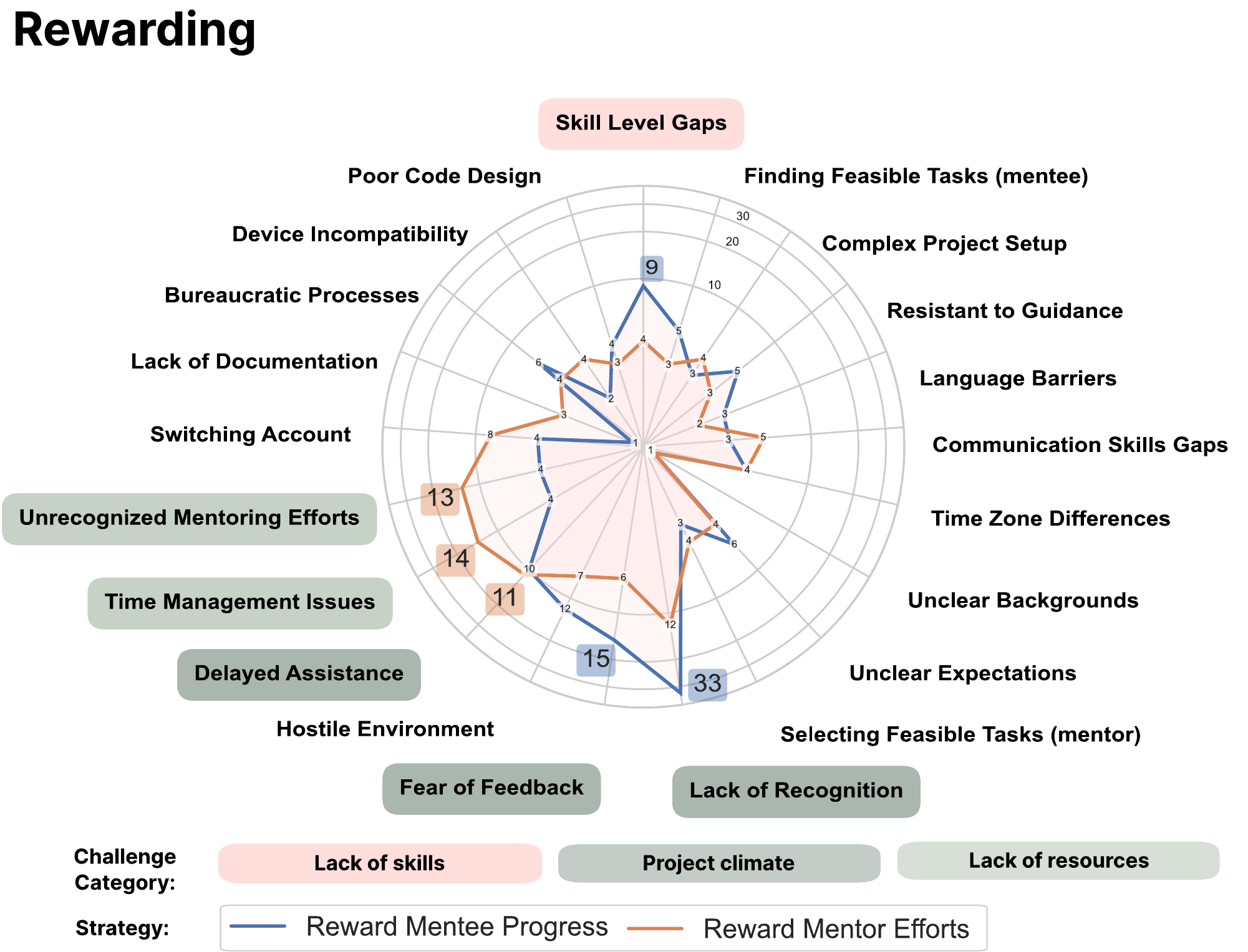}
\caption{Mapping of challenges to the strategy category \textbf{Rewarding}. The colored polygons represent strategies for \ul{rewarding mentees} (blue polygon) and \ul{mentors} (orange polygon). See Figure \ref{fig:preparing} for color-coding and layout details.} 
\label{fig:rewarding}
\vspace{-4mm}
\end{figure}

Figure \ref{fig:rewarding} show that mentors in our survey strongly agreed that \ul{rewarding} can address project climate challenges and lack of resources.


\textbf{Recognition in Promoting Positive Engagement} According to Self-Determination Theory \cite{deci2012self}, recognition and rewards are important for fulfilling psychological needs, such as competence, autonomy, and relatedness.  When \ul{mentees receive recognition} (blue polygon in Figure \ref{fig:rewarding}) for their contributions, it validates their efforts and helps improve their sense of competence, encouraging engagement. OSS mentors have mentioned that recognition and acknowledgment for mentees can reduce anxiety and increase motivation \cite{balali2018newcomers}. Ultimately, recognizing and rewarding mentees/mentors' efforts potentially create an inclusive, collaborative environment by fostering an appreciative and collaborative culture  \cite{feng2022case}.

\textbf{Rewarding Mentors to help manage time constraints } \ul{Rewarding mentors}, in turn, alleviates time management pressures by fostering intrinsic motivation to continue dedicating time to mentorship despite busy schedules \cite{ryan2000intrinsic}. When mentors feel their efforts are valued, this acknowledgment reinforces their commitment to mentoring, making them more inclined to dedicate time and energy to supporting others \cite{feng2022case}.

\begin{bubble}[Rewarding mentors and mentees...]
boosts engagement, supporting more supportive and collaborative communities.
\end{bubble}

\begin{figure}[!tbp]
\centering
\includegraphics[width=3.4 in]{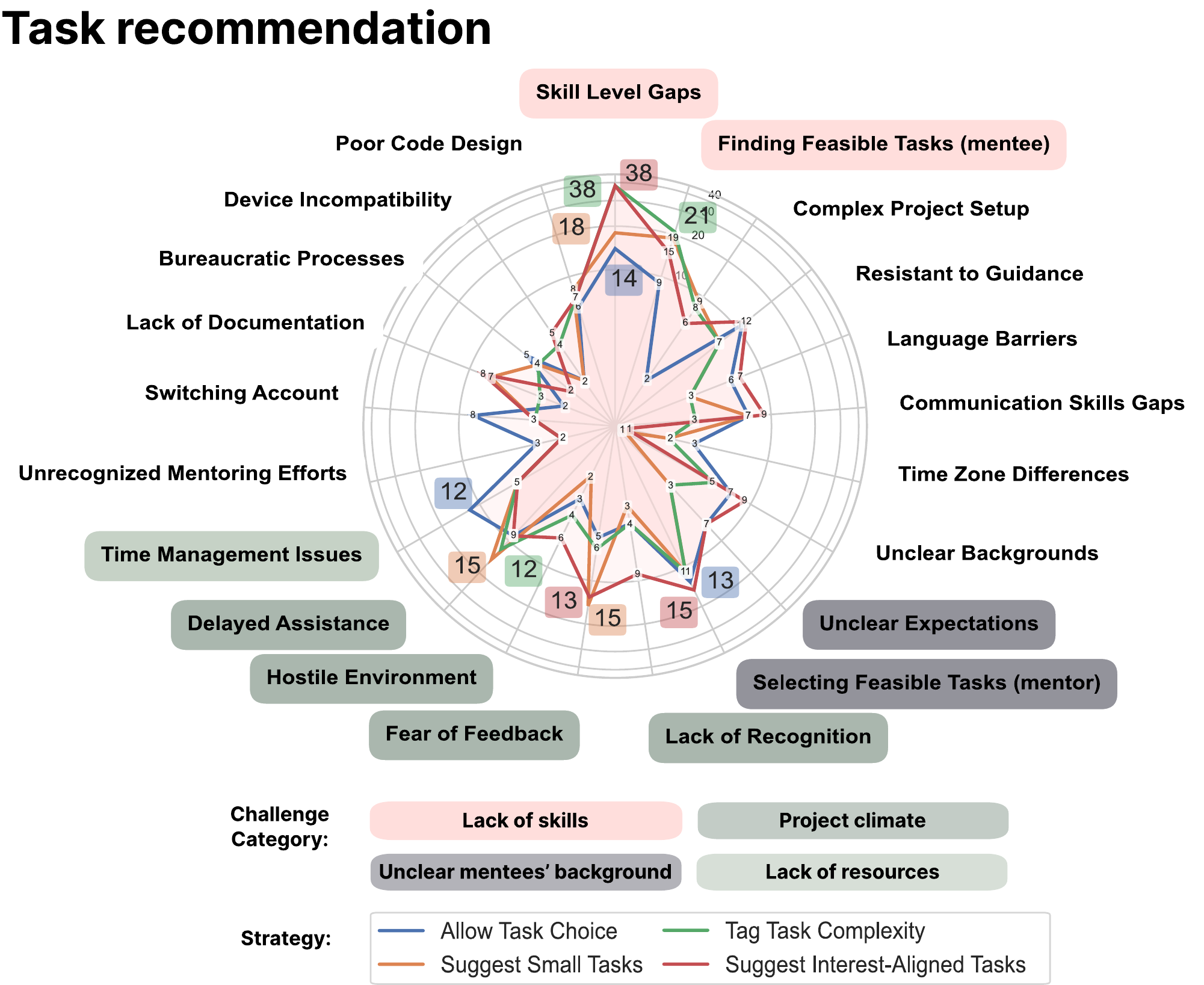}
\caption{Mapping of challenges to the strategy category \textbf{task recommendations}. The colored polygons in the spider plot represent strategies: \ul{allowing task choice} (blue polygon), \ul{tagging task complexity} (green polygon), \ul{suggesting small tasks} (orange polygon), and \ul{suggesting interest-aligned tasks} (red polygon). See Figure \ref{fig:preparing} for color-coding and layout details.} 
\label{fig:task}
\vspace{-4mm}
\end{figure}

\subsubsection{Task Recommendations} address challenges related to lack of skills, project climate, and unclear mentees' backgrounds (Figure \ref{fig:task}).

\textbf{Building Competence and Confidence} The use of strategies in the category of task recommendation is supported by theories in psychology and education that emphasize the importance of scaffolding in learning \cite{verenikina2008scaffolding}. Mentees can develop skills when guided through tasks that \ul{match their abilities and interests} \cite{verenikina2008scaffolding}. By \ul{suggesting small} (orange polygon in Figure \ref{fig:task}), \ul{repetitive tasks}, or \ul{tagging tasks by complexity} (green polygon), mentors provide mentees with a structured way to gradually build their competence and confidence to mitigate the challenges in skill level gaps, selecting feasible tasks.

\textbf{Reducing Dependence} In psychology, learners are more likely to exhibit autonomous problem-solving when provided with appropriately scaffolded tasks \cite{verenikina2008scaffolding}. When tasks are clearly defined, \ul{mentees can select tasks} (blue polygon) or be \ul{suggested tasks that match their skills and interests} (red polygon), reducing the need for constant guidance and fostering independent problem-solving skills. This, in turn, helps mentors mitigate time management challenges.

\begin{bubble}[Recommend tasks...]
such as suggesting small tasks and tagging complexity, helping mentees choose suitable tasks, fostering autonomy, and reducing mentor dependency.
\end{bubble}

\subsubsection{Establish Governance} 
\begin{figure}[!tbp]
\centering
\includegraphics[width=3.4 in]{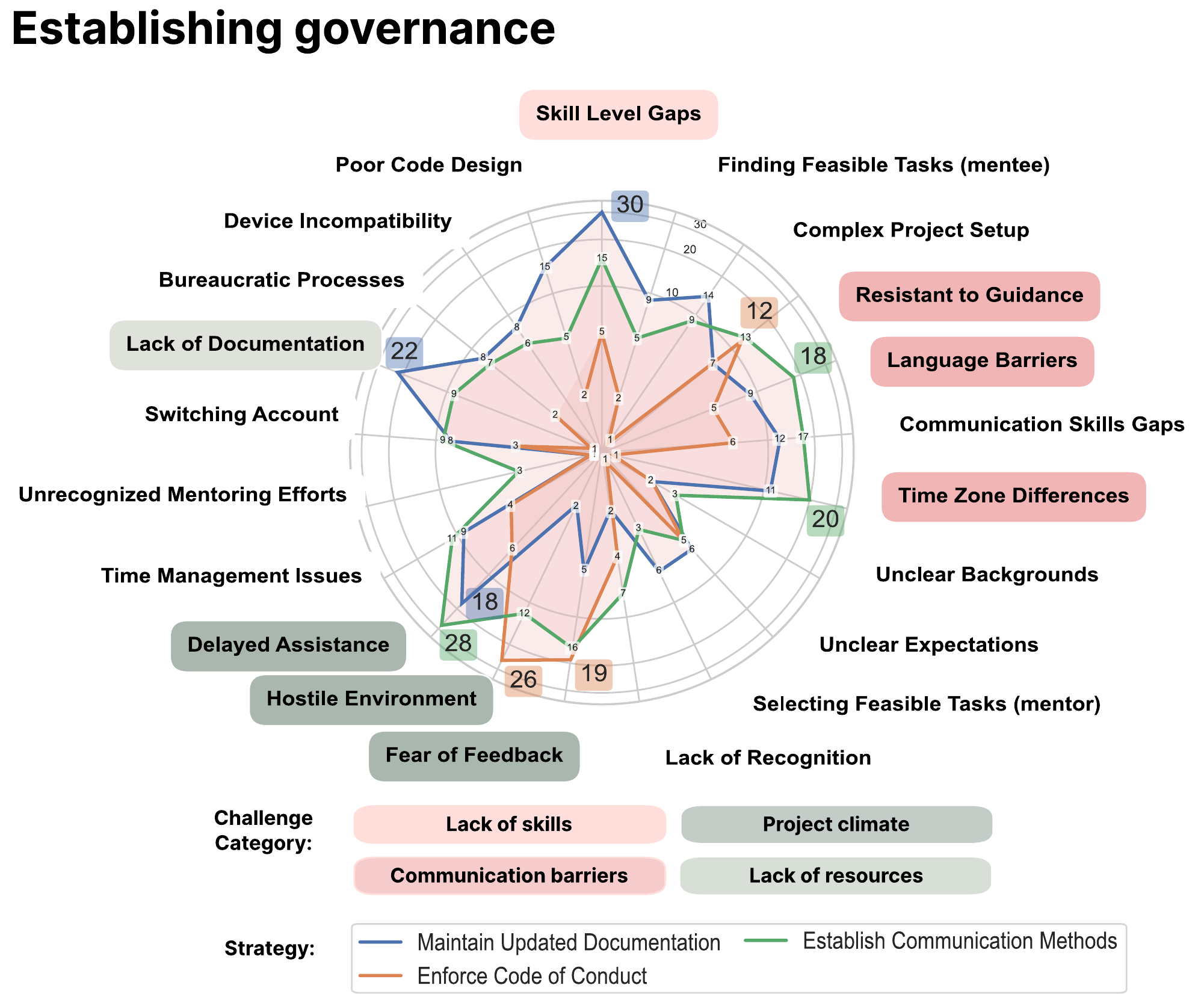}
\caption{Mapping of challenges to the strategy category \textbf{establishing governance}. The colored polygon in the spider plot represents strategies: \ul{maintain updated documentation} (blue polygon), \ul{establish communication methods} (green polygon), and \ul{enforce a code of conduct} (orange polygon). See Figure \ref{fig:preparing} for color-coding and layout details.} 
\label{fig:governance}
\end{figure}

As shown in Figure~\ref{fig:governance}, mentors in our survey indicated that strategies within the \ul{Establishing Governance} category focus on addressing challenges related to Project Climate. For example, the \ul{Code of Conduct} strategy (orange polygon) was reported to mitigate challenges such as mentees hesitating to submit code due to fear of harsh feedback, struggles with a hostile environment, and communication difficulties with mentees who resist guidance.

\textbf{Promoting Supportive Environments:} In OSS, toxic interactions are not uncommon; victims often become afraid to express themselves, which can lead to demotivation and, eventually, leave the project \cite{sarker2023automated}. Enforcing the code of conduct helps mitigate project climate challenges, such as a hostile environment, by managing potentially toxic conversations\cite{li2021code}.

\textbf{Facilitating Flexible Communication:} In OSS, with contributors worldwide, the relationship between mentees and mentors is often remote, facing challenges such as language barriers and timezone differences \cite{balali2018newcomers}. \ul{Establish various communication methods} and \ul{up-to-date documentation} in different languages help mentees receive timely support.

\begin{bubble}[Establish governance...]
with up-to-date documentation, codes of conduct, and flexible communication methods to tackle challenges related to skill gaps, project climate, communication barriers, and resource shortages.
\end{bubble}

\begin{figure}[!tbp]
\centering
\includegraphics[width=3.4 in]{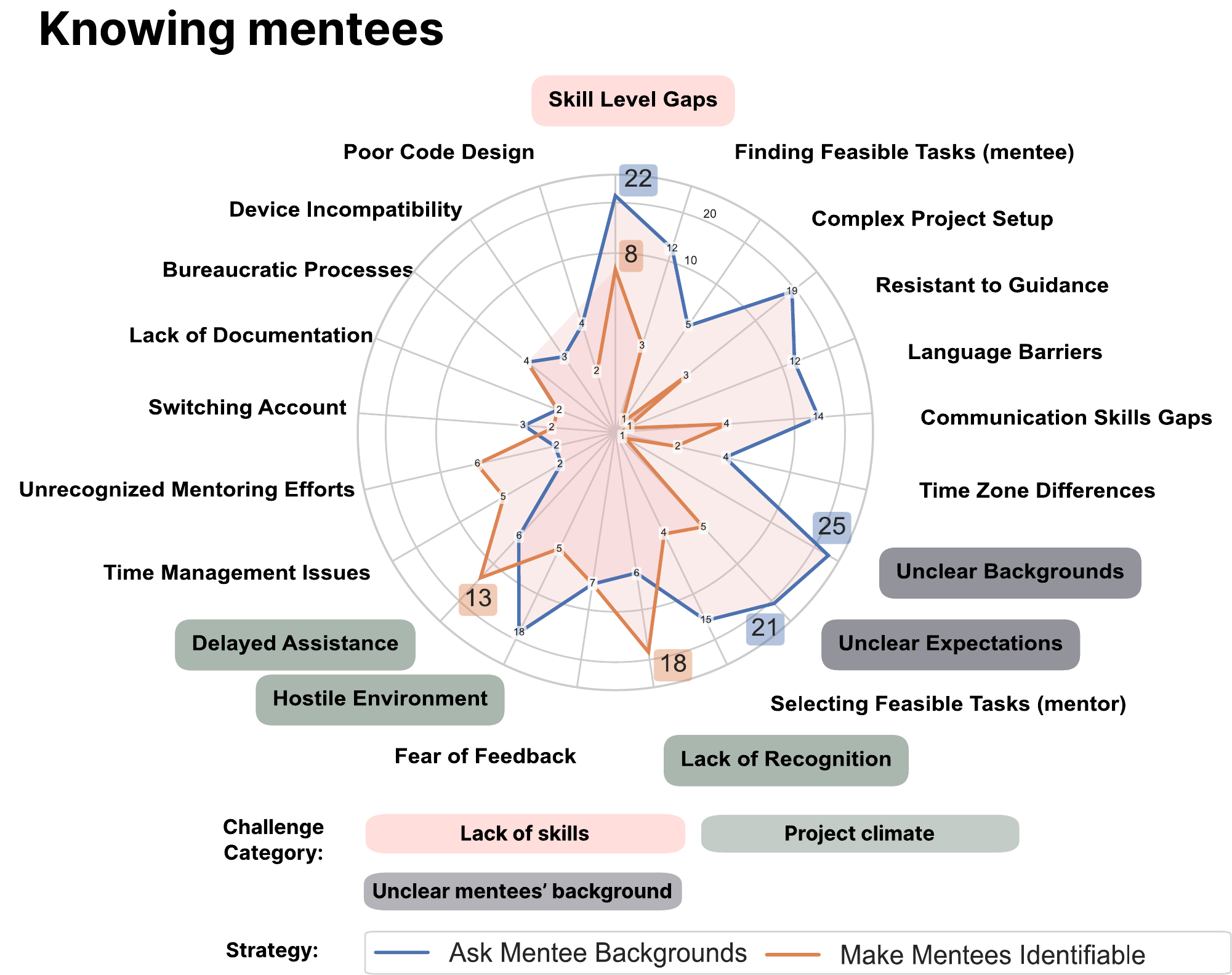}
\caption{Mapping of challenges to the strategy category \textbf{Knowing mentees}. The colored polygon in the spider plot represents strategies: \ul{ask mentee background} (blue polygon) and \ul{make mentees identifiable} (orange polygon). See Figure \ref{fig:preparing} for color-coding and layout details.} 
\label{fig:knowing_mentees}
\vspace{-3mm}
\end{figure}

\subsubsection{Knowing Mentees} includes two strategies for helping mentors know mentees: \ul{asking mentees background} and \ul{making mentees identifiable}: \textbf{Mentoring through mentee background awareness and visibility:} By actively understanding about the mentees' background and expectations (blue polygon in Figure \ref{fig:knowing_mentees}), mentors are better equipped to tailor their guidance, ensuring tasks align with mentees' abilities and helping to create a more personalized, supportive mentoring experience \cite{verenikina2008scaffolding}. When mentees are identifiable in the workspace, other community members are more likely to recognize their contributions and offer additional support, patience, and constructive feedback. This can help mitigate challenges such as delayed assistance, hostile environment, and lack of recognition.

\begin{bubble}[Knowing mentees...]
backgrounds and making them visible foster a sense of belonging and promote a collaborative environment.
\end{bubble}

\subsubsection{Supporting Mentees}  
\begin{figure}[!tbp]
\centering
\includegraphics[width=3.4 in]{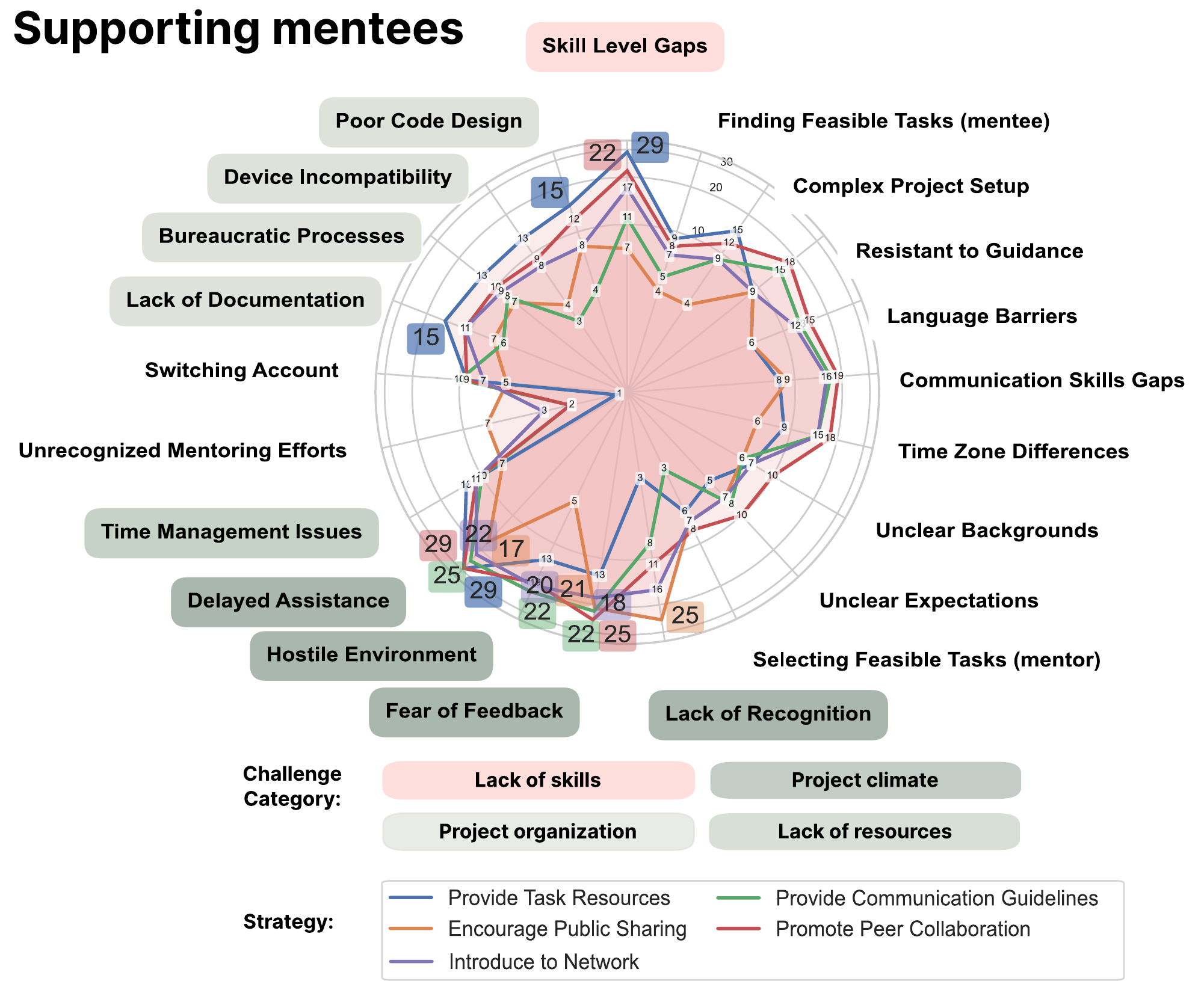}
\caption{Mapping of challenges to the strategy category \textbf{Supporting mentees}. The colored polygon in the spider plot represents strategies: \ul{provide task resources} (blue polygon), \ul{provide communication guidelines} (green polygon), \ul{encourage public sharing} (orange polygon), \ul{promote peer collaboration} (red polygon), and \ul{introduce to the network} (purple polygon). See Figure \ref{fig:preparing} for color-coding and layout details.} 
\label{fig:supporting_mentees}
\vspace{-3mm}
\end{figure}

Mentors in our survey (Figure \ref{fig:supporting_mentees}) think that strategies in \ul{Supporting Mentees} category address challenges related to \ul{Project Climate}, such as \ul{Project Organization (technical)} and \ul{Lack of Resources}.

\textbf{Creating Transparency and Confidence} By \ul{encouraging the sharing of work publicly} (orange polygon in Figure \ref{fig:supporting_mentees}), mentees are gradually encouraged to feel more comfortable contributing and receiving feedback. Similarly, \ul{providing communication guidelines} (green polygon) establishes clear expectations for interactions, helping to mitigate challenges such as hostile work environments and delayed responses \cite{feng2024guiding}.

\textbf{Building Social Support} \ul{Promoting peer collaborations} (red polygon) and \ul{introducing mentees to network} (purple polygon) also play a crucial role in supporting mentees. Mentoring is not limited to formal programs; implicit mentoring \cite{feng2022case};  within a community of practice, where mentees collaborate with peers, helps build social support networks and reinforce learning. These peer collaborations not only help mentees build their technical competence but also help reduce isolation in projects.

\textbf{Providing Task Resources} Learners benefit from resources (blue polygon) that enable them to solve problems by themselves, ultimately building their confidence and competence \cite{zimmerman2002becoming}. Strategies such as \ul{providing task resources} are reported to mitigate challenges beyond skill level gaps. This includes addressing issues like poor code design, complex project setups, and lack of documentation.

\begin{bubble}[Supporting mentees...]
 public sharing, clear communication, promoting peer collaboration, and offering task resources to bridge skill gaps, resolve recognition issues, and transform complex project setups into learning opportunities.
\end{bubble}

\textbf{Open-ended responses.}
We collected seven valid responses to the open-ended question about additional strategies participants would like to add. One mentor suggested \textit{``Periodic 1:1 meetings, having a co-mentor"} [P4]. Another two mentors suggested a foundation-based mentoring program to provide resource support for mentees, such as \textit{``Paid internships"} [P15], and provide \textit{``funding for mentees for things like laptops for the project"} [P8]. One participant discussed strategies for working with complex code design, \textit{``If the project is too complicated, or it’s a struggle to find feasible tasks for mentees; then mentors should work with the project on its architecture to reduce complexity and increase maintainability and modularity''} [P65]. 

Two participants suggested strategies for progress tracking and documentation \textit{``Provide a project tracking system to track progress and link resources''} [P40], \textit{``CI/CD and virtualization/containerization of build/development environments''} [P60]. Another participant emphasized challenges arising from the dynamics of OSS and highlighted the importance of building trust and empathy in mentoring relationships: \textit{``Most of these are interpersonal challenges that can only be addressed by close working relationships''} [P46].


\subsection{\SM}
\label{sec:result2}

\begin{figure}[!tbp]
\centering
\includegraphics[width=3.4in]{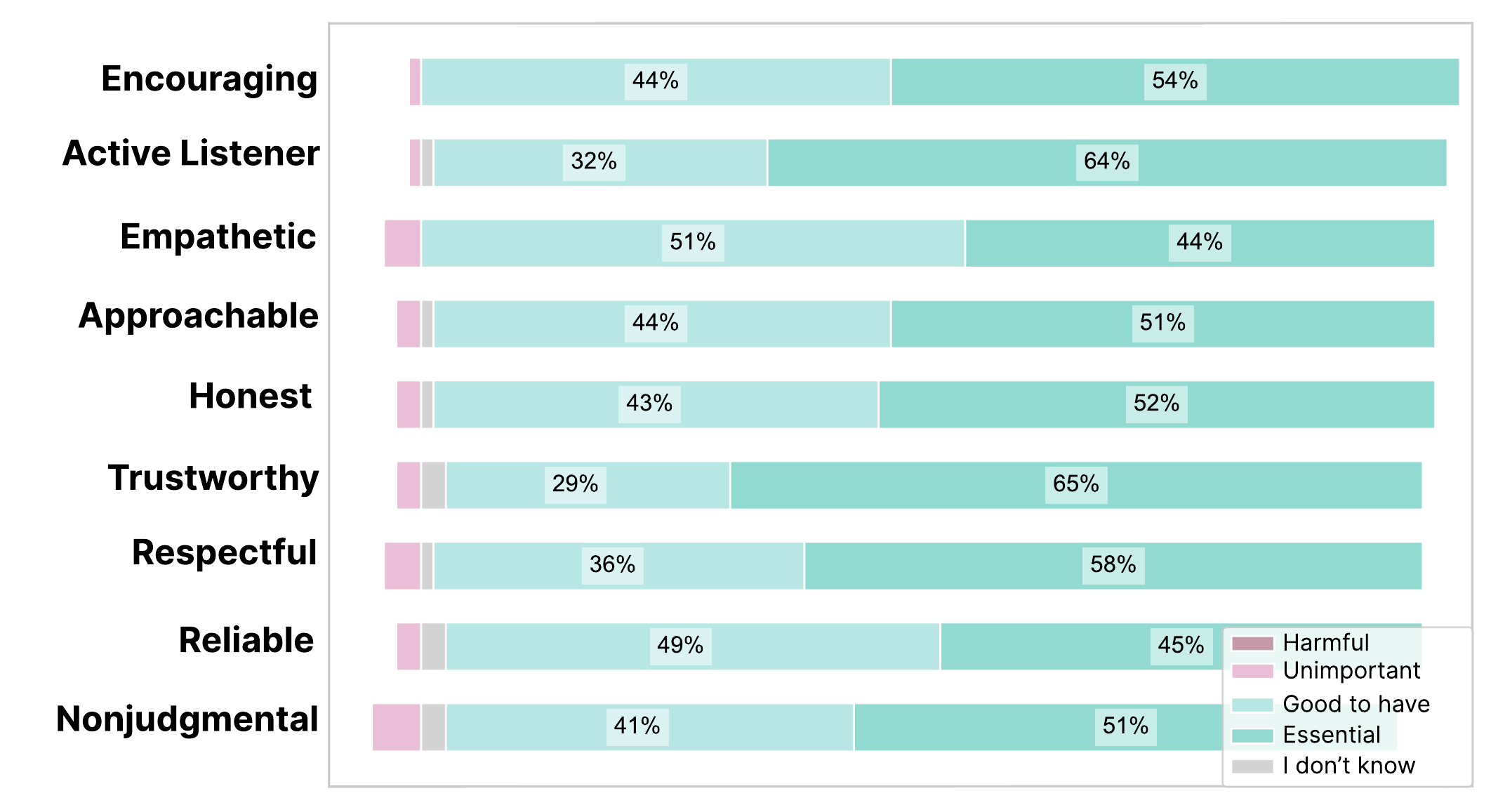}
\caption{Distribution of responses on the personal qualities of an ideal mentor (N=85). The bar plot displays the percentage of respondents.} 
\label{fig:quality}
\vspace{-4mm}
\end{figure}

Next, we focus on \SM{} that complement the technical mentoring strategies.

Figure \ref{fig:quality} shows the distribution of responses regarding the personal qualities of an ideal mentor (N=85). All qualities received strong agreement, with over 90\%  of either ``Good to Have'' or ``Essential''. Notably, \ul{Trustworthy}, \ul{Respectful}, and \ul{Active Listener}, were recognized by over 60\% of respondents as essential attributes for mentors. In the following sections, we discuss why these personal qualities are perceived as important attributes of good mentorship.

\textbf{Encouragement Fosters Belonging} The quality of \ul{being encouraging} was the highest-responsed quality of an ideal mentor, with 98\% of respondents rating it as either essential or good to have. In education, Tuckman et al. \cite{tuckman1991effect} concluded that encouragement fosters students' self-efficacy. By providing positive feedback, mentors can gradually motivate mentees to overcome challenges and sustain the mentoring relationship \cite{johnson2002intentional}.In OSS, challenges such as fear of harsh feedback are common \cite{sarker2023automated}, encouraging mentors can help mentees build confidence by recognizing their efforts \cite{feng2024guiding, zhang2011network}.

\textbf{The Value of Active Listening} \ul{Being an active listener} is the second most valued personal quality for an ideal mentor, with 64\% of respondents rating it as essential and 32\% as good to have. While mentors often excel at problem-solving and giving directions, it can be challenging for them to step back, listen, and ask questions \cite{ragins2007handbook}. Our findings suggest that active listening is equally important in OSS communities. For example, strategies in the category of knowing mentees, as discussed in Section \ref{sec:result1}, require mentors to actively listen to understand mentees' backgrounds and mitigate challenges such as skill level gaps and unclear expectations.

\textbf{The Importance of Empathy and Approachability} Two other highly valued qualities of ideal mentors are \ul{Empathetic} and \ul{Approachable}—both received over 95\% of responses (``Good to Have'' and ``Essential''). Studies have found that online environments can foster aggression compared to face-to-face interactions \cite{lapidot2012effects}, and toxicity remains a prevalent issue in OSS \cite{sarker2023automated}. When mentors show empathy, they can respond in ways that make mentees feel supported and understood. Empathetic mentors can offer tailored guidance, helping mentees navigate technical challenges and emotional stresses \cite{sarma2024effective}. Moreover, challenges such as hostile environments, where mentees may hesitate to ask questions or share their contributions, are common. Our findings in Section \ref{sec:result1} indicate that strategies in supporting mentees effectively mitigate these issues. However, such strategies are effective only when mentors are approachable, helping to build stronger relationships and encourage contributors' commitment to OSS. Lastly, our findings validate that \ul{Trust}, \ul{Honesty}, \ul{Respect}, and \ul{Reliability}—long recognized as key attributes in successful mentoring relationships—are equally important in OSS.

\begin{bubble}[Personal qualities...]
such as encouragement and active listening, fostering inclusion, and promoting contributor commitment.
\end{bubble}

\begin{figure}[!tbp]
\centering
\includegraphics[width=3.4in]{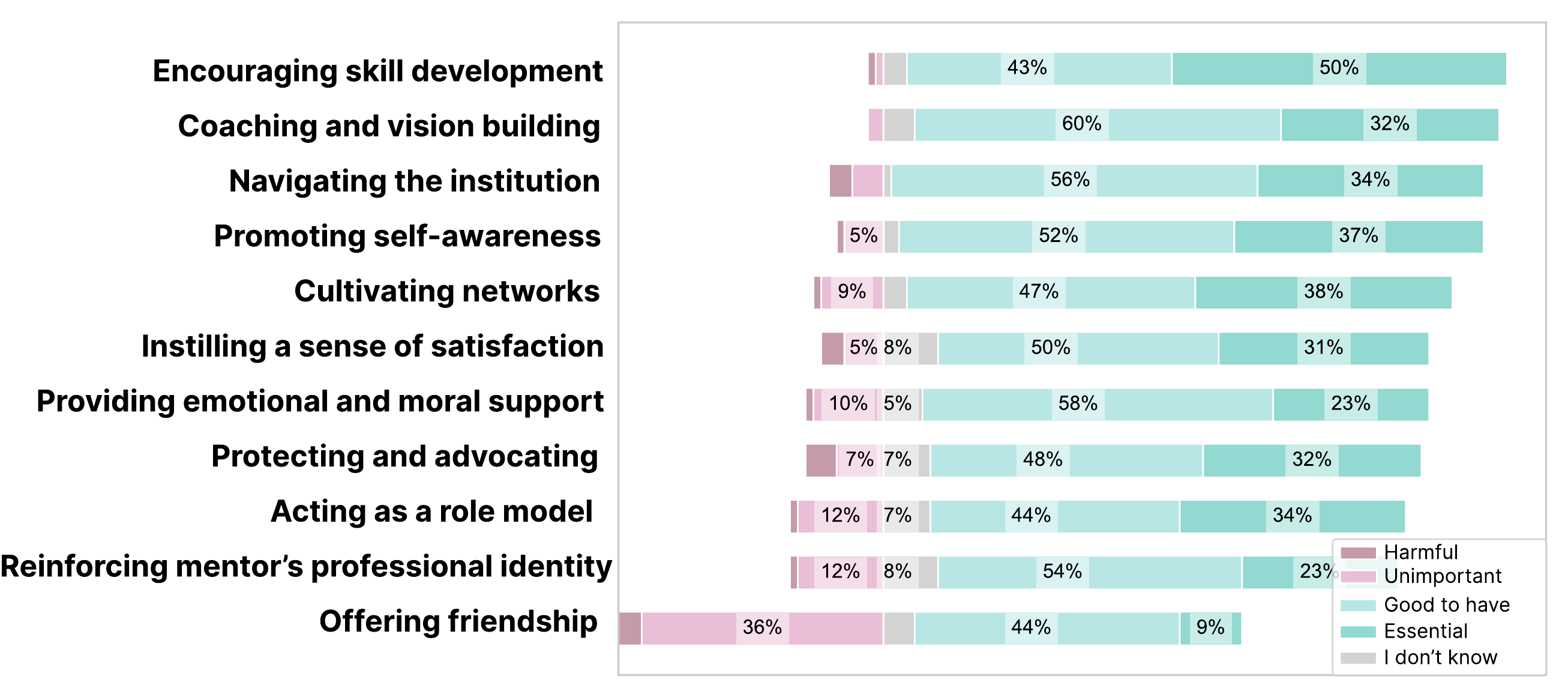}
\caption{Distribution of responses regarding the ideal mentoring outcomes (N=85). } 
\label{fig:outcome}
\vspace{-4mm}

\end{figure}

Next, we explore the ideal outcomes of mentorship. As shown in Figure \ref{fig:outcome}, \ul{Encouraging skill development} and \ul{coaching and vision building} received more than 90\% agreement of their importance to good mentorship (N=85).

\textbf{Technical Guidance Over Emotional Support:}
We have observed three outcomes that received more than 10\% of responses, marking them as unimportant, including \ul{Providing emotional and moral support}, \ul{Acting as a Role Model}, and \ul{Reinforcing the mentor's professional identity}. This contrasts with traditional mentoring models, where these outcomes are often viewed as crucial to the success of mentoring relationships \cite{picariello2016importance}. One possible reason is that many participants encounter goal-oriented mentoring arrangements.  For instance, internships at software companies often involve limited mentoring periods, such as summer internship programs \cite{microsoftInternship}. Similarly, in some OSS mentoring programs, the mentorship duration is typically brief—often around three weeks \cite{gsoc2024, menezes2022open}. Therefore, contributors seek technical guidance or solutions to immediate challenges rather than emotional or moral support.

\begin{bubble}[Mentorship in short-term/goal-oriented contexts]
often prioritizes technical guidance over emotional support due to time constraints, remote collaboration, and the need for immediate solutions.
\end{bubble}

\textbf{Professional Boundaries of Mentoring:} Moreover, nearly half of the participants did not agree that friendship is an essential or desirable outcome of mentorship. This is understandable, as remote work is increasingly the norm in software development \cite{ford2021tale}. Similarly, in OSS mentoring programs, where contributors often collaborate across cultures, regions, and time zones, it can be challenging to build friendships \cite{feng2024guiding}. A study found that emotional entanglements in professional mentoring relationships can reduce the mentor’s ability to provide objective and constructive feedback, as they may hesitate to criticize the work of a friend \cite{ragins2007handbook}. Therefore, it is possible that many participants viewed OSS mentorship as a professional relationship focused on skill development, knowledge sharing, and career growth and not as a personal bond that might blur professional boundaries.

\begin{bubble}[Friendship is not agreed...]
as an ideal outcome of mentorship by our participants, as it may blur professional boundaries and hinder objective feedback in a diverse, asynchronous environment.
\end{bubble}
\label{sec_results}
\vspace{-3mm}

\section{Discussion}

In this section, we summarize implications based on our findings for mentoring as a call to action to help mentors, mentees, and organizations.

\subsection{How our findings are useful for practitioners}

\textbf{Practical Strategies from a Technical Perspective}: The results from the \TM{} survey provide actionable insights for mentors in OSS, especially new mentors, as these strategies are informed by the perspectives of experienced mentors who have worked with various mentees. Equipped with a toolkit of these strategies, mentors can support mentees through challenges and help facilitate task completion.

\textbf{Need for Social Support}: The \SM{} survey results offer mentors an understanding of good mentorship beyond technical task completion. By providing a deeper understanding of personal qualities and expected outcomes associated with good mentorship, these insights ultimately promote both mentee and mentor confidence and long-term engagement in OSS. Participants in our study rated emotional support as of lower importance. Future research should investigate if emotional support may be undervalued and how mentorship programs can be designed to address mentees’ holistic needs, thereby creating a more inclusive and supportive OSS community.

\textbf{Pre-Screening for Mentor Recruitment}: Organizations like CodeDay and the Google Summer of Code (GSoC) could use the \SM{} survey findings to design pre-screening surveys for recruiting mentors. These organizations can recruit mentors who match the ideal qualities of a mentor and the expected outcomes.

\subsection{Opportunities for future work}

\textbf{Towards an Evidence-Based Handbook for Effective OSS Mentorship}: Based on the findings from our \TM{}, \SM{} survey results, we plan to develop a handbook for effective mentorship practices in OSS. This handbook could be shared among mentors, providing them with insights gained from experienced mentors. For instance, when a mentor or mentee is hesitant to adopt strategies like starting with small or repetitive tasks, the organization can refer to the study’s findings to emphasize the perceived effectiveness of these approaches.

\textbf{Designing an AI Mentor For OSS/SE}
We believe future research could leverage strategies to design AI-driven interventions, such as chatbots integrated with large language models \cite{kasneci2023chatgpt}. By incorporating the strategies from the \TM{}, the chatbot can suggest technical strategies, using tone and intended outcomes as identified through the \SM{} survey findings. This would reduce the load on mentors, who are already resource-constrained, and provide mentoring to newcomers tailored to their (technical) needs and (social) expectations.

\label{sec_discussion}
\vspace{-2mm}

\section{Threats to Validity}

As with all empirical studies, ours carries risks to validity; we argue that we have taken reasonable steps to mitigate the effects of these potential threats. In this section, we describe these mitigation steps in detail.

\textbf{Internal}. The wording and comprehension of survey questions may lead to misunderstandings. To reduce this threat, we piloted both surveys in multiple rounds with different groups of pilot participants and revised the survey incrementally based on their feedback until no further comments were received.

\textbf{External}. We acknowledge that the findings may not be generalized to all OSS communities. To reduce these threats, we avoided focusing on a single community; instead, we used a multi-approach survey distribution to reach a broadly representative sample.

Another threat to this study is the sample size of the surveys. Finding participants for this study was nontrivial, as it required mentors who were already overloaded to take extra time to share their experiences. We employed multiple distribution methods to increase participation. We avoided reaching out through OSS mailing lists, such as GitHub user accounts, as this could be considered spam under community codes of conduct and might raise ethical concerns \cite{baltes2016worse}. We think the sample size for both studies is reasonable to draw conclusions from, as the count of participants aligns with the literature on anthropology, suggesting that a minimum of 10 knowledgeable participants is sufficient to discover and understand the core categories in studies of lived experiences \cite{bernard2017research}. Furthermore, our number of participants exceeds similar mapping studies in OSS \cite{balali2020recommending}.

\textbf{Conclusion}. Our study consists of two surveys with different participant samples. However, we note that many participants are from North America. We acknowledge that such regional skew limits the generalizability of our findings. In the future, when we acquire more data, more broadening recruitment strategies should be considered.

\vspace{-2mm}
\section{Conclusion and Future Work}

In this study, we first investigated strategies that contributors perceive as effective for addressing task-related challenges in mentorship. We also analyzed the personal attributes of ideal mentors and ideal mentorship outcomes that can help create effective mentor-mentee pairings. In future work, we plan to conduct longitudinal studies within a mentoring program, such as the CodeDay mentoring program, to evaluate the effectiveness of these strategies in practice.

\vspace{-2mm}
\section{Acknowledgment}
We thank the Outreachy group for their valuable feedback on the survey design and our survey respondents. We also thank Diane Mueller and Nate Waddington for their contributions. This work is partially supported by NSF grants 1901031, 2235601, 2236198, 2247929, 2303042, 2303043, and 2347311.

\label{sec_conclusion_future}
\vspace{-3mm}

\begingroup
\tiny
\bibliographystyle{IEEEtran}
\bibliography{bib.bib}

\bibliographystylelit{IEEEtran}
\bibliographylit{bib_lit}
\endgroup

\end{document}